\documentclass[superscriptaddress,showpacs,aps]{revtex4-1}
\usepackage{amsmath}
\usepackage{graphicx}
\begin{document}

\title{\bf Exact relativistic models of perfect fluid disks in a  magnetic field}

%formato prd

\author{Gonzalo Garc\'{\i}a-Reyes}
\email[e-mail: ]{ggarcia@utp.edu.co}
\affiliation{Departamento de F\'{\i}sica, Universidad Tecnol\'ogica de Pereira,
 A. A. 97, Pereira, Colombia}

 \author{Omar A. Espitia}
\email[e-mail: ]{omar.espitia@upb.edu.co}
\affiliation{Departamento de Ciencias B\'asicas, Universidad Pontificia Bolivariana,
 Bucaramanga, Colombia}

\begin{abstract}
Using the well-known ``displace, cut and reflect'' method we construct
thin disks made of a perfect fluid  in presence of a magnetic field.
The models are based in a  magnetic Reissner-Nordstrom metric
of  Einstein-Maxwell equations for a conformastatic spacetime.
The influence of the magnetic field on the matter properties of the disk are  analyzed.
We  also study  the motion of  charged  test particles   around  the disks.
We construct  models of  perfect fluid disks satisfying all  the energy conditions.
\end{abstract}

\maketitle

\section{Introduction}

Axially symmetric exact solutions of Einstein equations describing the field  of a thin disk are
important in two context, in astrophysics as models of certain stars, galaxies, accretion disks,
and the superposition of a black hole and a galaxy or an accretion disk as
in the case of quasars,
and in general relativity as sources
of exact solutions of Einstein field's equations.
Thin disks  in presence of magnetic field  
are also of astrophysical importance  in the
study of active galactic nuclei, neutron stars, white dwarfs and galaxy formation,
or  as sources of Einstein-Maxwell equations.
For example, in the case of  highly magnetized neutron stars, or magnetars,
these fields are characterized by a  strength of the order of  
$10^{14}$ G \cite{neu1, neu2, neu3}.
To describe the gravitational and electromagnetic fields
of such configurations, is necessary  to study the coupled Einstein-Maxwell
equations.

Exact solutions of the Einstein equations representing the
field of a static thin disks without radial pressure were first studied by  Bonnor and Sackfield \cite{BS},
and Morgan and Morgan \cite{MM1}, and with radial  pressure also by Morgan and Morgan
\cite{MM2}.  
Several classes of exact solutions of the  Einstein field  equations
corresponding to static thin disks with or  without radial pressure have been
obtained by different authors
\cite{LP,CHGS,LO,LEM,BLK,BLP,GE}.
Rotating  thin disks that can be considered as a source of a Kerr metric were
presented by  Bi\u{c}\'ak and  Ledvinka \cite{BL}, while rotating disks with
heat flow were were studied by Gonz\'alez and Letelier \cite{GL2}.
The exact superposition of a disk and a static black hole was first considered
by Lemos and Letelier in Refs. \cite{LL1,LL2,LV-AGN}. 
%and perfect fluid disks with haloes in Ref. \cite{VL-halos}.
On the other hand, thin disks in presence
of electromagnetic field have been discussed as sources for 
Kerr-Newman fields  \cite{LBZ},
conformastationary metrics \cite{KBL},
and magnetostatic  axisymmetric fields  \cite{LET1,GG1,GG2,GG-CRIS,AN-GUI}. 
Perfect fluid disks as sources of Taub-NUT-type spacetimes were presented
in Ref. \cite{GG3}, and  static charged perfect fluid disks  were   study in \cite{VL}.
Also, conformastatic disk-haloes in Einstein-Maxwell gravity were considered in \cite{AN-GUI-QUE},
variational thermodynamics of relativistic thin disks
in \cite{AN-QUE},
and the exact superposition of a disk and a static black hole in a magnetic field in  
\cite{AN-GON-GUI}.

In this work we consider exact relativistic models of thin disks made of perfect fluid
in presence of a magnetic field.
The models are constructed using the well-known ``displace, cut and reflect''
method. These disks are essentially  of infinite extension. However,  since  the surface energy density
of the disks decreases rapidly  one can define a cut off radius, of the order of the  galactic disk radius,  and, in principle, to
consider these disks as finite. Moreover,   even though realistic
disklike sources have thickness, in first approximation these astrophysical objects can
be considered to be very thin, e.g., in our Galaxy the radius of the disk is 10 kpc
and its thickness is 1 kpc. The disks also satisfy  all the energy condition, which means that these
structures are made of usual matter and no tachyonic or otherwise exotic matter.

The paper is organized as follows. In Sec. II we present
the formalism  to construct   models of  thin disks  made of a perfect fluid
and  electric current for a conformastic spacetime.
We also analysis the motion of  charged  test particles   around  the disks.
In Sec. IV  a simple family  of  perfect fluid disks in presence of a magnetic field
is considered based in a magnetic Reissner-Nordstrom metric
of  Einstein-Maxwell equations for a conformastatic
spacetime.  Finally, in Sec. V we summarize and discuss the results obtained.

%---------------------------------------------------------------------
%---------------------------------------------------------------------

\section{Einstein-Maxwell equations and perfect fluid disks  }

The metric   for a conformastatic  spacetime  can be written as  \cite{AN-GUI-QUE}
\begin{equation}
ds^2 =  -e^{2 \psi } dt^2 + e^{ 2 \lambda } (R^2 d \varphi ^2 + d R^2 + dz^2  ),
\label{eq:metconf}
\end{equation}
where $(\varphi, R,z)$ are the  usual cylindrical coordinates.
For  axially symmetric  fields
$\psi$ and $\lambda$  are functions of the coordinates $R$ and $z$
only. The vacuum Einstein-Maxwell equations, in  geometrized units such that
$ G = c  = 1$,  are given by
\begin{subequations}\begin{eqnarray}
&   &    R_{ab} \  =  \ 8 \pi T_{ab}, \label{eq:einmax1}  \\
&   &   \nabla_b F^{ab} = 0,  \label{eq:einmax2}
\end{eqnarray}\end{subequations}
where
\begin{equation}
T_{ab} \  =  \ \frac{1}{4 \pi} \left [ F_{ac}F_b^{ \ c} - \frac 14
g_{ab}F_{cd}F^{cd} \right ] \label{eq:tab}
\end{equation}
is the electromagnetic energy-momentum  tensor,
\begin{equation}
F_{ab} =  A_{b,a} -  A_{a,b}
\end{equation}
the electromagnetic field tensor, and $A_a$  the  electromagnetic four
potential.
The other symbols have the usual meaning, i.e.,  $( \ )_{,a}=\partial /\partial
x^a$,  $\nabla_b$   covariant derivate, etc.

Solutions of the Einstein-Maxwell equations (\ref{eq:einmax1}) - (\ref
{eq:einmax2})  representing the field of a thin disk at $z=0$ with electric
current
can be constructed assuming the components of the metric tensor and
the electromagnetic potential  continuous across the disk, and its first
derivatives discontinuous in the direction normal to the  disk.  This can be
written  as
\begin{subequations}\begin{eqnarray}
b_{ab} \ &=& [g_{ab,z}] =  g_{ab,z}|_{_{z = 0^+}} \ - \ g_{ab,z}|_{_{z = 0^-}}
 \ = \ 2 \ g_{ab,z}|_{_{z = 0^+}}, \label{eq:b}  \\
 a_{b} \ &=&  [A_{b,z}] =  A_{b,z}|_{_{z = 0^+}} \ - \ A_{b,z}|_{_{z = 0^-}}
 \ = \ 2 \ A_{b,z}|_{_{z = 0^+}}.    \label{eq:a}
\end{eqnarray}\end{subequations}

The application of the formalism of distributions in curved spacetimes to
the Einstein-Maxwell equations \cite{PH,LICH,TAUB,IS1,IS2} give us
\begin{subequations}\begin{eqnarray}
R_{ab} & = &  8 \pi T_{ab} , \label{eq:EMdisk}  \\
T_{ab} & = &  T^{\text{elm}}_{ab} + T^{\text{mat}}_{ab}
= T^{\text{elm}}_{ab} +  Q_{ab} \ \delta (z) , \label{eq:Tabdisk}  \\
 \nabla_b F^{ab} & = & 4 \pi J^a ,  \\
J^a & = & j^a \delta (z) ,
\end{eqnarray}\end{subequations}
where $\delta(z)$ is the  usual Dirac function with support on  the disk,
$T^{\mathrm {elm}}_{ab}$ is the electromagnetic tensor (\ref{eq:tab}),
\begin{equation}
Q^a_b = \frac{1}{16 \pi}\{b^{az}\delta^z_b - b^{zz}\delta^a_b +  g^{az}b^z_b -
g^{zz}b^a_b + b^c_c (g^{zz}\delta^a_b - g^{az}\delta^z_b)\}  \label{eq:Qab}
\end{equation}
is the  energy-momentum tensor on plane $z=0$, and
\begin{equation}
j^a =  \frac{1}{4 \pi} [F^{ab} ] \delta^z_b, \label{eq:ja}
\end{equation}
is the electric current density on the disk. $[F^{ab} ]$ means the jump of
Maxwell tensor across the disk.
The ``true''  surface
energy-momentum tensor (SEMT) of the  disk $S_{ab}$ and the ``true'' surface
current density $\text{\sl j}_a$ are given by
\begin{subequations} \begin {eqnarray}
S_{ab} & = & \int T^{\mathrm {mat}}_{ab} \ ds_n =  \sqrt{g_{zz}} Q_{ab}  ,   \label{eq:Sab}  \\
\text{\sl j}_a  & = & \int J_{a}  \ ds_n = \sqrt{g_{zz}} j_a ,  \label{eq:jatrue}
\end{eqnarray}\end{subequations}
where $ds_n = \sqrt{g_{zz}} \ dz$ is the ``physical  measure'' of length in the
direction normal to the disk.  For the metric (\ref{eq:metconf}), the nonzero
components of  $S_a^b$ are
\begin{subequations}\begin{eqnarray}
&S^t_t &= \ \frac{1}{2 \pi} e^{-  \lambda}  \lambda_{,z} ,
 \label{eq:emt1}     		\\
&	&	\nonumber	\\
&   S^\varphi_\varphi  &=    S^R_R  = \ \frac{1}{4 \pi} e^{- \lambda} (\psi + \lambda)_{,z}  . \label{eq:emt2}
\end{eqnarray}\label{eq:emt}\end{subequations}
For axially symmetric  magnetostatic fields the  magnetic potential is given by  $A_a=\delta_a ^\varphi A$
where $A$ is functions of $R$ and $z$ only.
The only  nonzero component of the current density is
\begin{equation}
\text{\sl j}_{\varphi} = \ - \frac{1}{2 \pi} e^{- \lambda}
A _{,z}.  \label{eq:jacimut}
\end{equation}
All the quantities are evaluated at $z = 0^+$.

In terms of the orthonormal  tetrad ${{\rm e}_{ (a)}}^b = \{
V^b , W^b , X^b,Y^b \}$, where
\begin{subequations}\begin{eqnarray}
V^a &=& e^{-  \psi}    \delta ^a_t  , \quad  	
W^a =  \ \ e^{-  \lambda }  \delta ^a_\varphi / R , \label{eq:tetrad1}	\\
X^a &=& e^{-   \lambda }  \delta ^a_R , \quad
Y^a = e^{- \lambda }   \delta ^a_z  , \label{eq:tetrad2}
\end{eqnarray}\end{subequations}
the SEMT can be written as
\begin{equation}
S^{ab} = p g^{ab} + (p + \epsilon )V^a V^b, \label{eq:perfect}
\end{equation}
where
\begin{equation}
\epsilon \ = \ - S^t_t , \quad  p \ =  \  p_\varphi \ = \ p_R  =  \ S^\varphi_\varphi ,
\label{eq:enpr}
\end{equation}
are, respectively,  the surface energy density $\epsilon$ and  the
azimuthal and radial  pressure $p$ on the disk.
The azimuthal current density \text{\sl j} is given by
\begin{equation}
\text{\sl j}  = W^\varphi \text{\sl j}_\varphi \label{eq:j}.
\end{equation}

Thus, these disks can be interpreted as a matter distribution made of a perfect fluid with electric current.

%movimiento de particulas

We will now analyze the motion of  charged  test particles   around the disks.
For equatorial circular orbits the 4-velocity  $u^a$  of the particles with respect to the coordinates
frame has  components $u^a = u^0(1,\omega, 0,0 )$ where  $\omega= u^1/u^0=\frac{d
\varphi}{d t}$ is  the  angular speed of the test particles.  For the spacetime  (\ref{eq:metconf}),
the equation for the electrogeodesic motion of the particle is given by
\begin{equation}
\frac 12   g_{ab, R}u^a u^b = - \tilde e  F_{R a} u^a_,
\label{eq:geo}
\end{equation}
where $\tilde e$ is the specific electric  charge of the particles.  In the case magnetostatic
the motion equation  reads
\begin{equation}
 \frac 12 u^0 (g_{\varphi \varphi, R } \omega ^2 + g_{t t, R} )= - \tilde e
A_{, R} \omega, \label{motion}
\end{equation}
where $u^0$ obtains normalizing  $u^a$, that is requiring  $g_{ab}u^au^b=-1$, so that
\begin{equation}
 (u^0)^2 = - \frac {1}{g_{\varphi \varphi} \omega^2 + g_{tt}} \label{eq:u0}.
\end{equation}
Thus, the angular speed  $\omega $  is given by
\begin{equation}
\omega ^2 = \frac { - T_1 \pm \sqrt { T_1^2-T_2 T_3 }}{ 2  T_2 }, \label{eq:omega}
\end{equation}
where
\begin{subequations}\begin{eqnarray}
T_1 & = & 2 g_{t t, R}  g_{\varphi \varphi, R} +  4 \tilde e ^2 g_{ t
t}  A_{, R}^2,    \\
T_2 & = &  g_{\varphi \varphi, R}^2  + 4 \tilde e ^2  g_{ \varphi \varphi}
A_{, R}^2 ,  \\
T_3 & =& 4 g_{t t, R}^2 .
\end{eqnarray}\end{subequations}

The  positive sign corresponds to the direct orbits or co-rotating
and the negative sign  to the retrograde orbits or counter-rotating.

With respect to the  orthonormal tetrad (\ref{eq:tetrad1})-(\ref{eq:tetrad2}) the 3-velocity
has components
\begin{equation}
 v^{ (i)}=  \frac { {{\rm e}^{ (i)}}_a u^a } { {{\rm e}^{(0)}}_b u^b } .
\end{equation}
For equatorial circular  orbits  the
only nonvanishing velocity components is given by
\begin{equation}
 (v^{ (\varphi)})^2 = v_c^2= - \frac{ g_{\varphi \varphi} }{ g_{tt} } \omega ^2  \label{eq:vc2}
\end{equation}
which represents   the circular speed   of the particle as seen by an observer at infinity.

%ejemplo simple

\section{Disks from a  magnetic Reissner-Nordstrom solution }

A simple exact solution of Maxwell-Equations for the  spacetime (\ref{eq:metconf}) is given by the metric
\begin{subequations}
\begin{eqnarray}
ds^2 & =& -   \frac{ \left (  1 - \frac {k^2} {4  r^2}   \right )^2}{ \left (  1  + \frac{\tilde a
+ \tilde b}{2  r}  \right )^2
\left (  1  + \frac{\tilde a - \tilde b}{2  r}    \right )^2  } dt^2
+ \left (  1  + \frac{\tilde a + \tilde b}{2  r}  \right )^2
\left (  1  + \frac{\tilde a - \tilde b}{2  r}    \right )^2 (R^2 d \varphi ^2 + d R^2 + dz^2 ),  \label{eq:metiso}  \\
A&=& \tilde b \left (  \frac{ z}{r } +1  \right),
\end{eqnarray}
\end{subequations}
where $r=\sqrt{R^2+ z^2}$, $\tilde a = k a$, $\tilde b = kb$, being $k$, $a$, and  $b$ constants.  $a^2=1+b^2$ and  $b$ is the parameter that controls
the magnetic field.  By written the solution in  spherical coordinates $(r, \theta, \varphi)$
and then  using  the radial coordinate transformation  \cite{VL}
\begin{equation}
 r' =  r   \left (  1  + \frac{\tilde a + \tilde b}{2  r}  \right )    \left (  1  + \frac{\tilde a
 - \tilde b}{2  r}    \right ),
\end{equation}
one find  that  this spacetime  is  the magnetic Reissner-Nordstrom metric
\begin{subequations}
\begin{eqnarray}
ds^2 & = & -   \left ( 1 - \frac{2 \tilde a}{r'}  + \frac{\tilde b^2}{r'^2}  \right ) dt^2
+ \left ( 1 - \frac{2 \tilde a}{r'}  + \frac{\tilde b^2}{r'^2}  \right ) ^{-1}  dr'^2 + r'^2 d \Omega ^2 ,
\label{eq:metsch} \\
A  &=& \tilde  b (\cos \theta +1),
\end{eqnarray}
\end{subequations}
where $d \Omega ^2 = d \theta ^2 + \sin ^2 \theta d \varphi ^2$.
The solution is the  magnetostatic counterpart of the  Reissner-Nordstrom solution
and can be obtained using the   well-know theorem proposed by Bonnor in  \cite{Bonnor}. It also
can be generated   using the Ernst method  \cite{E1,E2,GG-CRIS} from   Schwarzschild
solution \cite{Sch}. Indeed for $b=0$, it goes over into the Schwarzschild
solution.

Exact solutions which represent the field of a  disk
can be obtained using  the well known ``displace,
cut and reflect'' method  that was first
used by Kuzmin \cite{KUZMIN} and Toomre \cite{TOOMRE} to constructed Newtonian
models of disks, and later extended to general relativity
\cite{BLK,BLP,BL,GL2}. Given a solution of the
Einstein-Maxwell equation, this procedure is mathematically equivalent to apply
the transformation $z \rightarrow |z| + z_0$,
with $z_0$ constant. The method is an adaptation of the method
of images of electrostatic.

From (\ref{eq:enpr})-(\ref{eq:j}), the main  physical quantities associated
with the system  are
\begin{subequations}
\begin{eqnarray}
\tilde \epsilon &=& \frac{ 4 \alpha ( 2 a \bar r  + 1   ) }
{ \pi (4 \bar r^2 +2a \bar r  + 1 )^2    },  \\
\tilde p &= & \frac{ 2 \alpha   ( 4 \bar r^2 + 4  a \bar r  +1   ) }
{  \pi  ( 4 \bar r ^2  - 1 ) (4 \bar r^2 +2a \bar r  + 1 )^2 },  \\
 \text{\sl j} &=& - \frac{ 8 b \bar r^3  } {  \pi (4 \bar r^2 +2a \bar r  + 1 )^2  }   ,
\end{eqnarray}
\end{subequations}
where $\tilde \epsilon = k \epsilon$, $\tilde p = k p$, and
$ \bar r = \sqrt{\tilde R^2 + \alpha^2} $, with $\tilde R = R /k$ and
$\alpha = z_0 /k$. Since $\epsilon \geq 0$  these  disks satisfy
the weak energy condition  for all values of parameter and we have always  pressure (stress positive) for $\alpha > 1/2$.
It follows that the strong energy condition $\epsilon + 2 p \geq 0 $ is also satisfied.
The dominant  energy condition requires  that $p \leq  \epsilon  $  which  for $\alpha > 1/2$ reads
$(2 \bar r ^2)( 8a \bar r) + 4 \bar r ^2  + 3 \geq  8a \bar r $. It follows that these disks  also satisfy this condition.

On the other hand, the lines of force of the magnetic field are given by the
equation $A(r,z)=C$, being $C$ a constant \cite{LET1}.
Calling $C=C_A$,  the lines satisfy the expression
\begin{equation}
 R = \frac{\sqrt{C_A (2kb-1)}}{C_A-kb} ( |z|+ z_0 ).
\end{equation}
Thus, to order to have  well defined  lines $kb \geq 1/ 2$ and  $C_A>kb$.
The lines of force of the gravitational field can be represent 
by drawing the equipotential lines which  are  given by $g_{00}=C$, being $C$ also a constant.
Gravitational field lines are perpendicular  to  equipotential  lines. 
Making $C=-C_G^2$, the equipotential  lines satisfy the relation
\begin{equation}
 R^2+ (|z| + z_0)^2 = \tilde C_G^2,
\end{equation}
where
\begin{equation}
 \tilde C_G = \frac {- a C_G \pm \sqrt{ a^2 C_G^2 + 4 k^2(1-C_G^2) }}{4(C_G-1)},
\end{equation}
with $-1\leq C_G <1$.

%Analisis graficas

\begin{figure}
$$
\begin{array}{ccc}
\tilde \epsilon    &  \tilde p &  -\mbox{\sl j} \\
\includegraphics[width=0.3\textwidth]{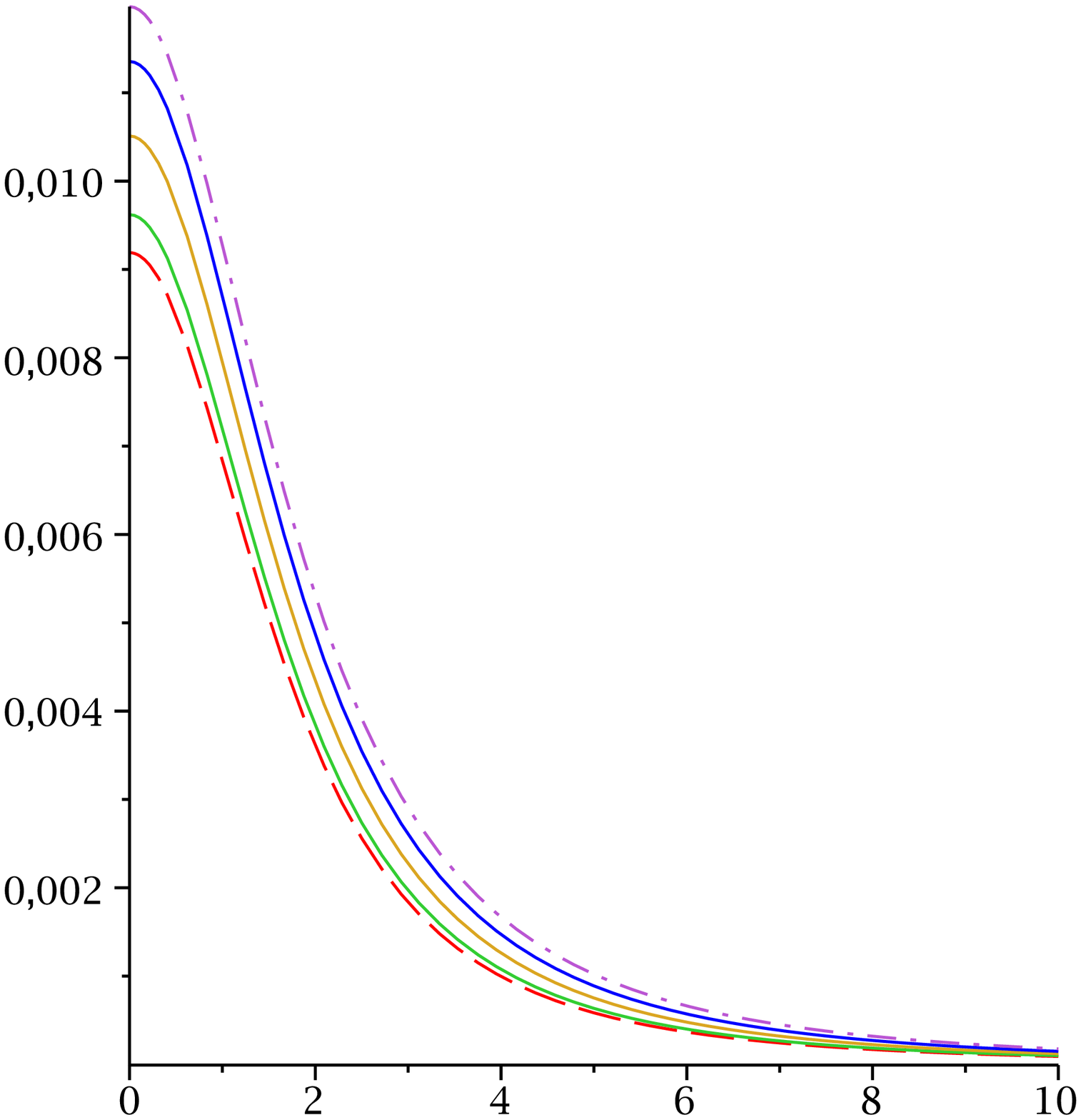} &
\includegraphics[width=0.3\textwidth]{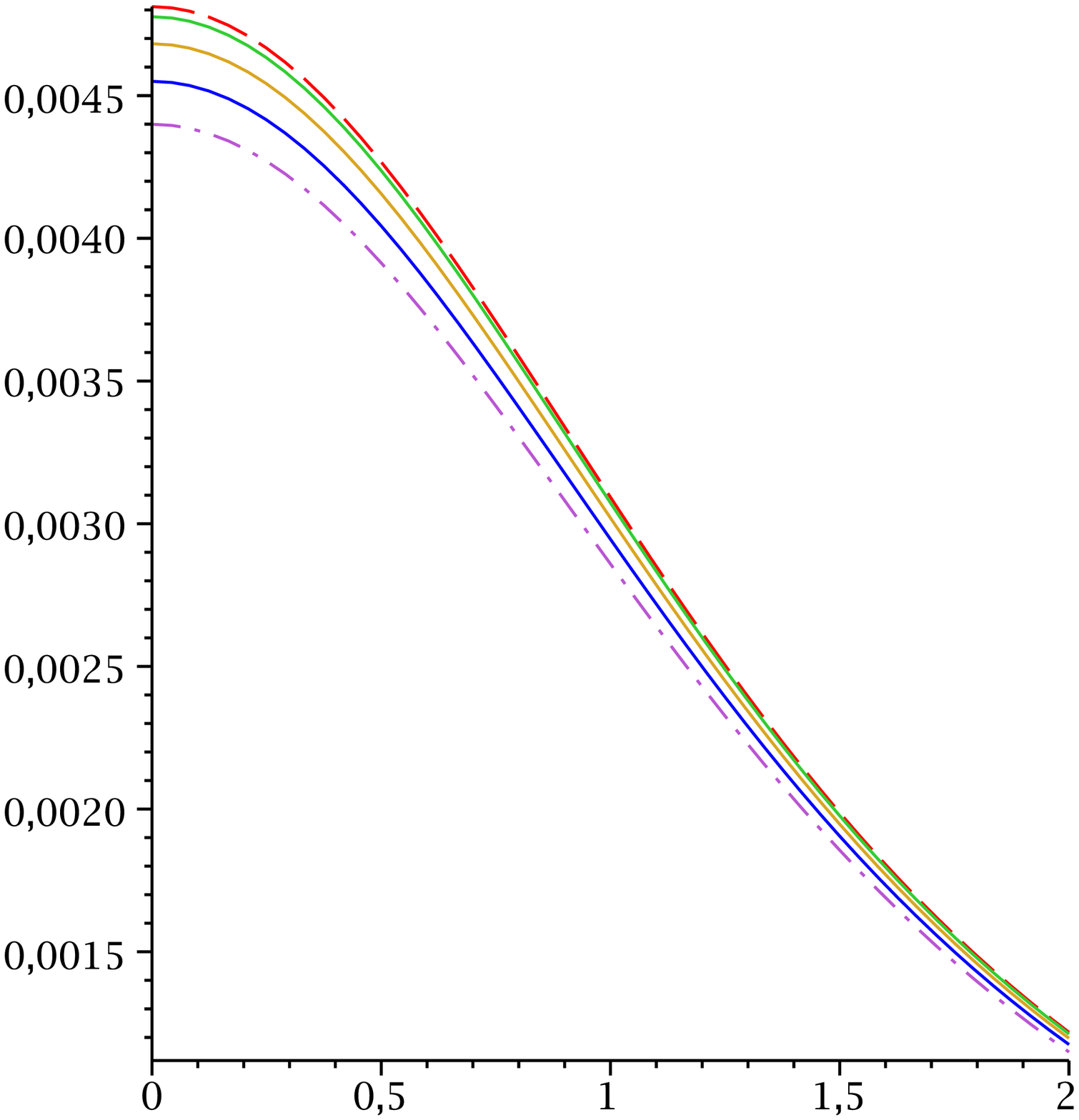} &
\includegraphics[width=0.3\textwidth]{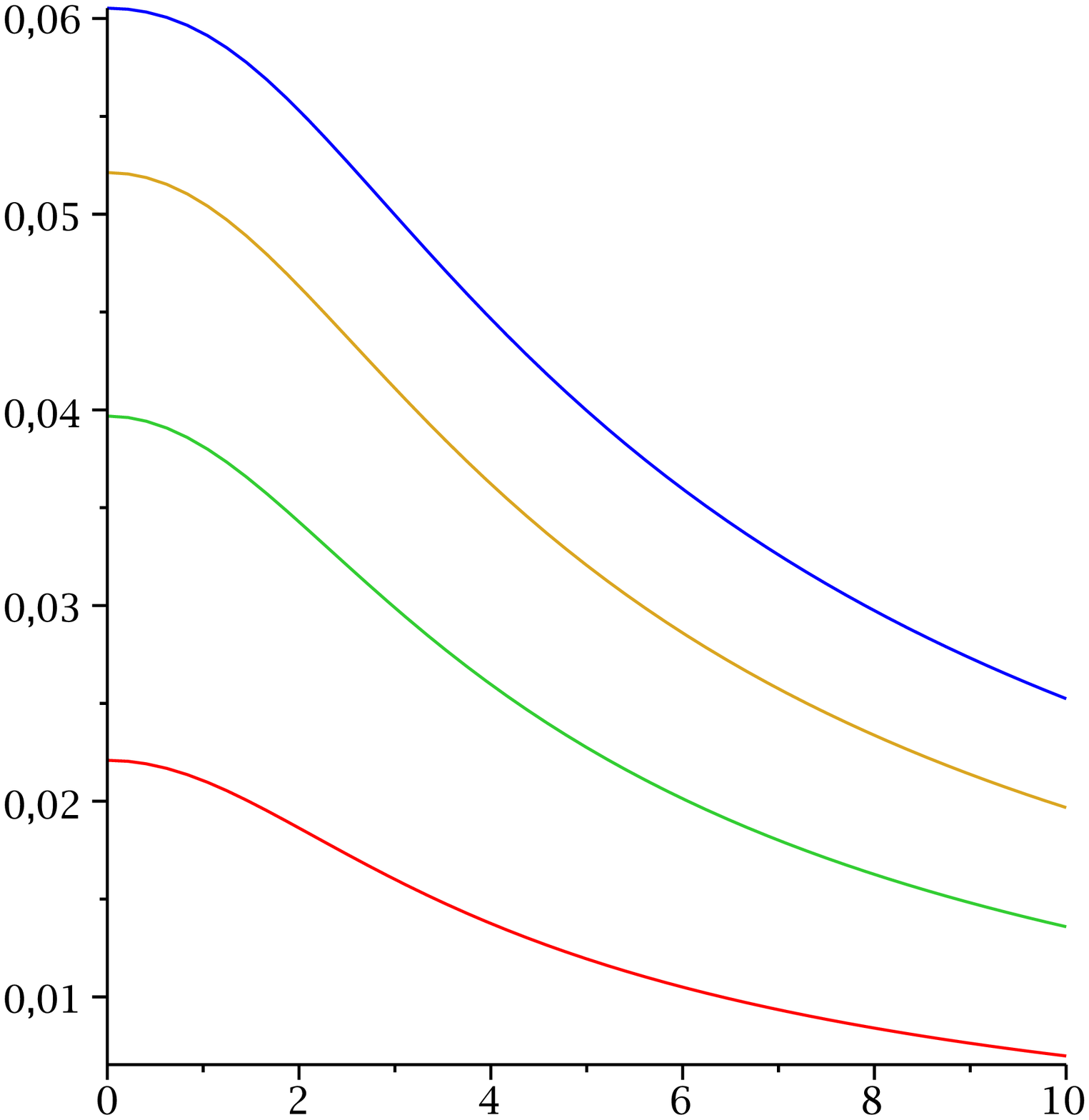}   \\
 \tilde R & \tilde R & \tilde R \\
(a)    &  (b) & (c)
\end{array}
$$	
\caption{$(a)$  The  energy density $\tilde \epsilon$ and   $(b)$ the
pressure $\tilde p$ for  perfect fluid disk in a magnetic field with  $\alpha=2$
and   values of magnetic field parameter  $b= 0$ (dashed curves), $0.5$,  $1$, $1.5$,
and $2$ (dash-dotted  curves), as  functions of $\tilde R$. $(c)$.  The azimuthal electric
current density $\mbox{\sl j}$ for
$b= 0$ (axis $\tilde R$), $0.5$,  $1$, $1.5$,  and $2$ (top curve) and the same value of
the parameter $\alpha$.}
\label{fig:fig1}
\end{figure}

In Figs. \ref{fig:fig1}$(a)$  -  \ref{fig:fig1}$(c)$, we show   the  surface energy density
$\tilde \epsilon$, the pressure $\tilde p$
and the azimuthal  electric current density $\text{\sl j}$  for disks  with cut parameter  $\alpha =2$ and  values of
the parameter of magnetic field
$b= 0$, $0.5$,  $1$, $1.5$, and $2$,  as  functions of $\tilde R$.  
We can see that the energy is everywhere
positive, its maximum occurs in the center of the
disk, and it vanishes sufficiently fast as r increases.
This permits  to  define a cut off radius and, in principle, to
model these structures as a compact object. 
We observe that the pressure and the electric current  have  a behavior similar to
the energy.
We find that
the  the presence of  magnetic field
increases  the energy density  and the electric current density  while   the
pressure is lowered everywhere on the disks. 
We also computed these functions for other
values of the parameters within the allowed range and in all
cases we found a similar behavior.

\begin{figure}
$$
\begin{array}{cc}
 v^2_+   &  v^2_-  \\
\includegraphics[width=0.3\textwidth]{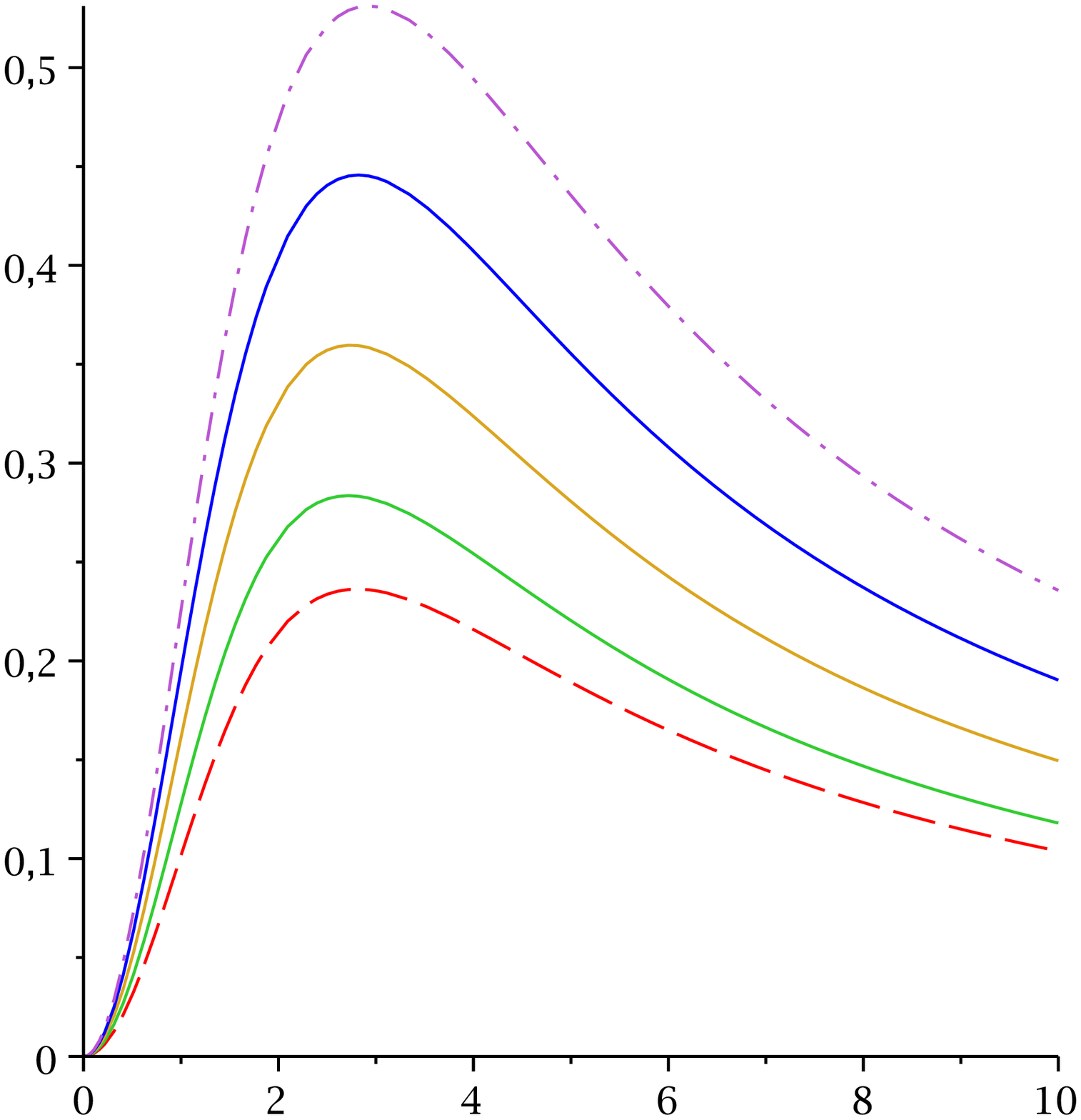} &
\includegraphics[width=0.3\textwidth]{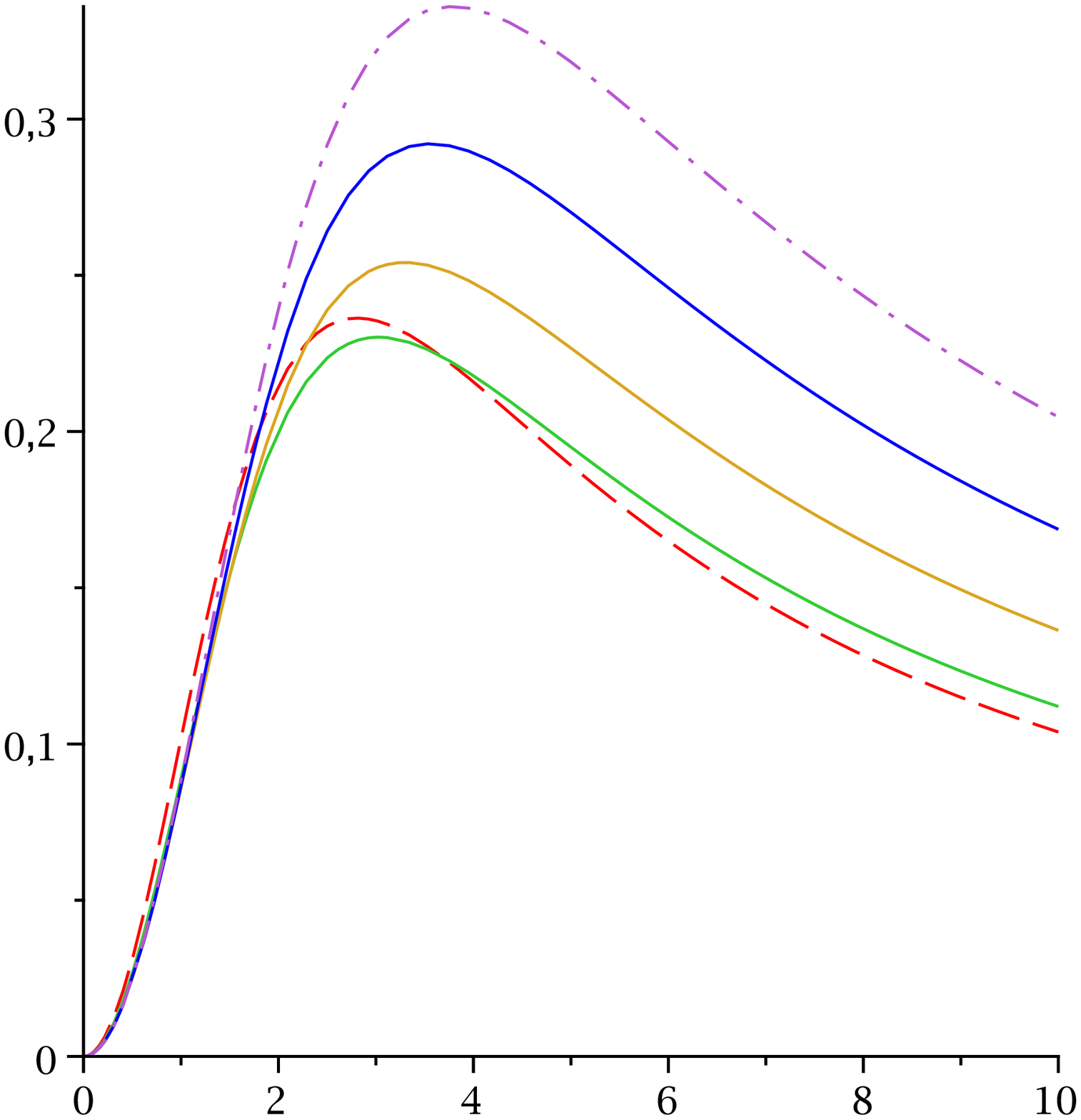} \\
\tilde R  &  \tilde R \\
(a)    &  (b)    \\
\includegraphics[width=0.3\textwidth]{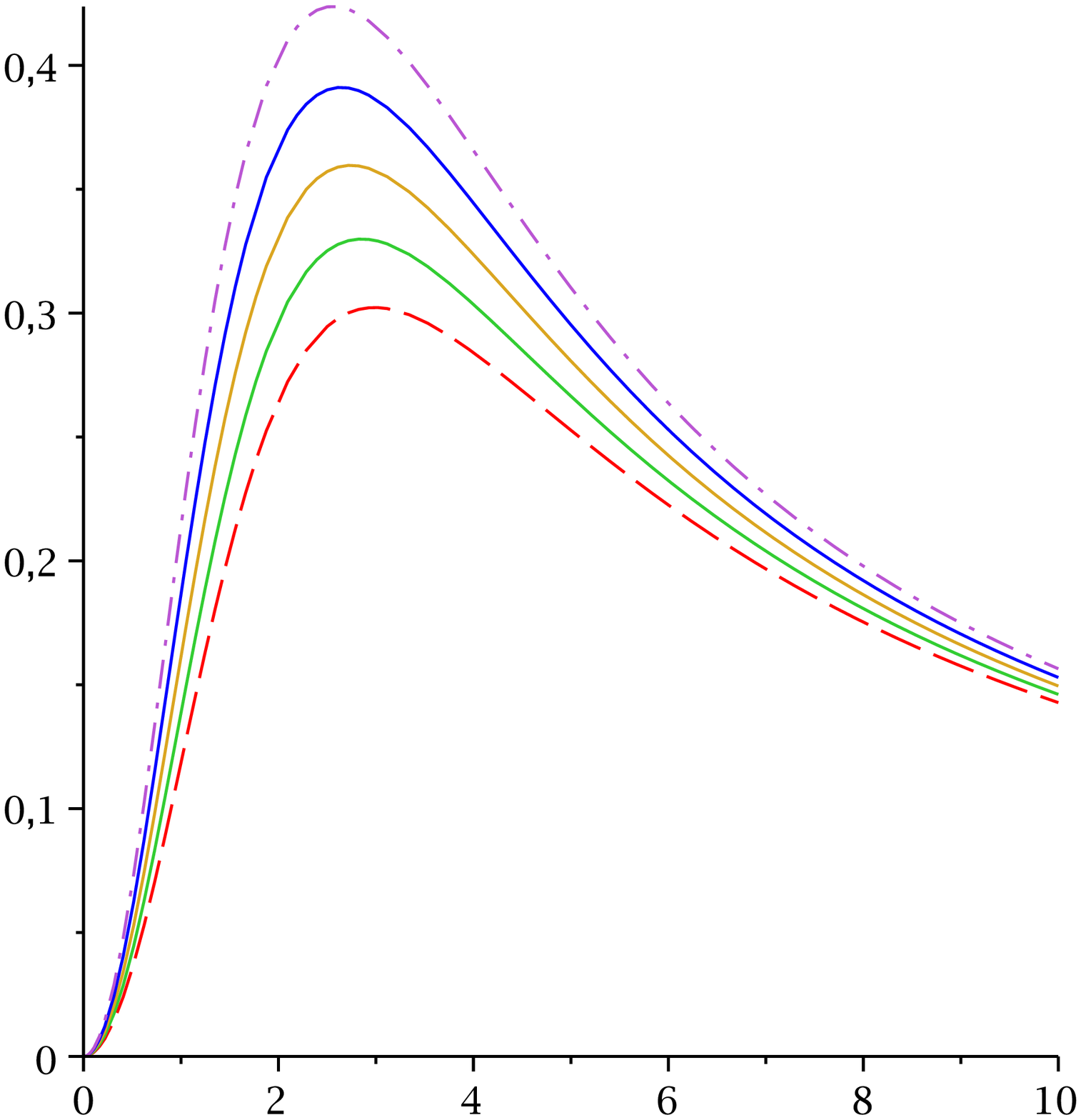} &
\includegraphics[width=0.3\textwidth]{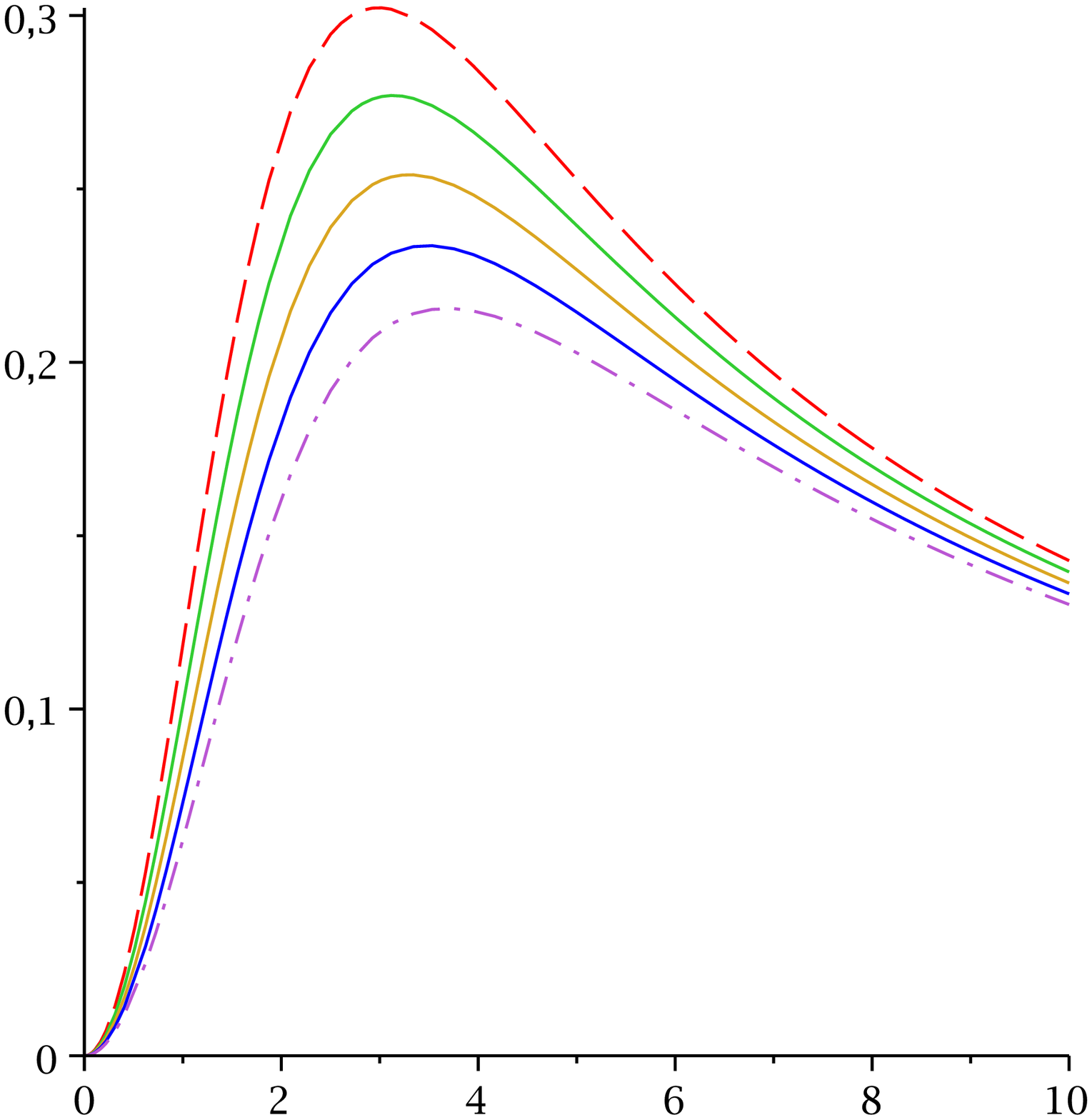} \\
\tilde R  &  \tilde R \\
(c)    &  (d) 
\end{array}
$$	
\caption{The circular  speed  $(a)$  $v^2_+$ and   $(b)$ $v^2_-$ 
for  for $\alpha=2$,  $\tilde e = 1$,   
$b= 0$ (dashed curves), $0.5$,  $1$, $1.5$,
and $2$ (dash-dotted  curves).
Again the circular  speed  $(c)$  $v^2_+$ and   $(d)$ $v^2_-$ for   the same value of  $\alpha$, $ b=1$,
$\tilde e= 0$ (dashed curves), $0.5$,  $1$, $1.5$,
and $2$ (dash-dotted  curves). }
\label{fig:fig2}
\end{figure}

In Figs. \ref{fig:fig2}$(a)$ and \ref{fig:fig2}$(b)$, we illustrate the behavior of the
circular speed for direct orbit $v^2_+$ and  retrograde orbits  $v^2_-$ of  charged test particles
with  $\tilde e = 1$,   $b= 0$ (dashed curves), $0.5$,  $1$, $1.5$,
and $2$ (dash-dotted  curves) and the same value of  $\alpha$.
We observer  that  circular speed  of particles is always a quantity
less than the speed  of light and that the magnetic field makes more relativistic these orbits.
In Figs. \ref{fig:fig2}$(c)$ and  \ref{fig:fig2}$(d)$, we also plot  the circular speed
for the same value of $\alpha$,  $ b=1$, and different values of the specific electric charge
$\tilde e= 0$ (dashed curves), $0.5$,  $1$, $1.5$,
and $2$ (dash-dotted  curves).
We find that for direct orbits  the specific charge  increases the speed of the particles whereas
for retrograde orbit the contrary occurs. We also find that for neutral particles $\tilde e=0$ the speed of
particles is the same  for both direct and indirect orbits, whereas for charged particles are
different.

\begin{figure}
$$
\begin{array}{cc}
\includegraphics[width=0.3\textwidth]{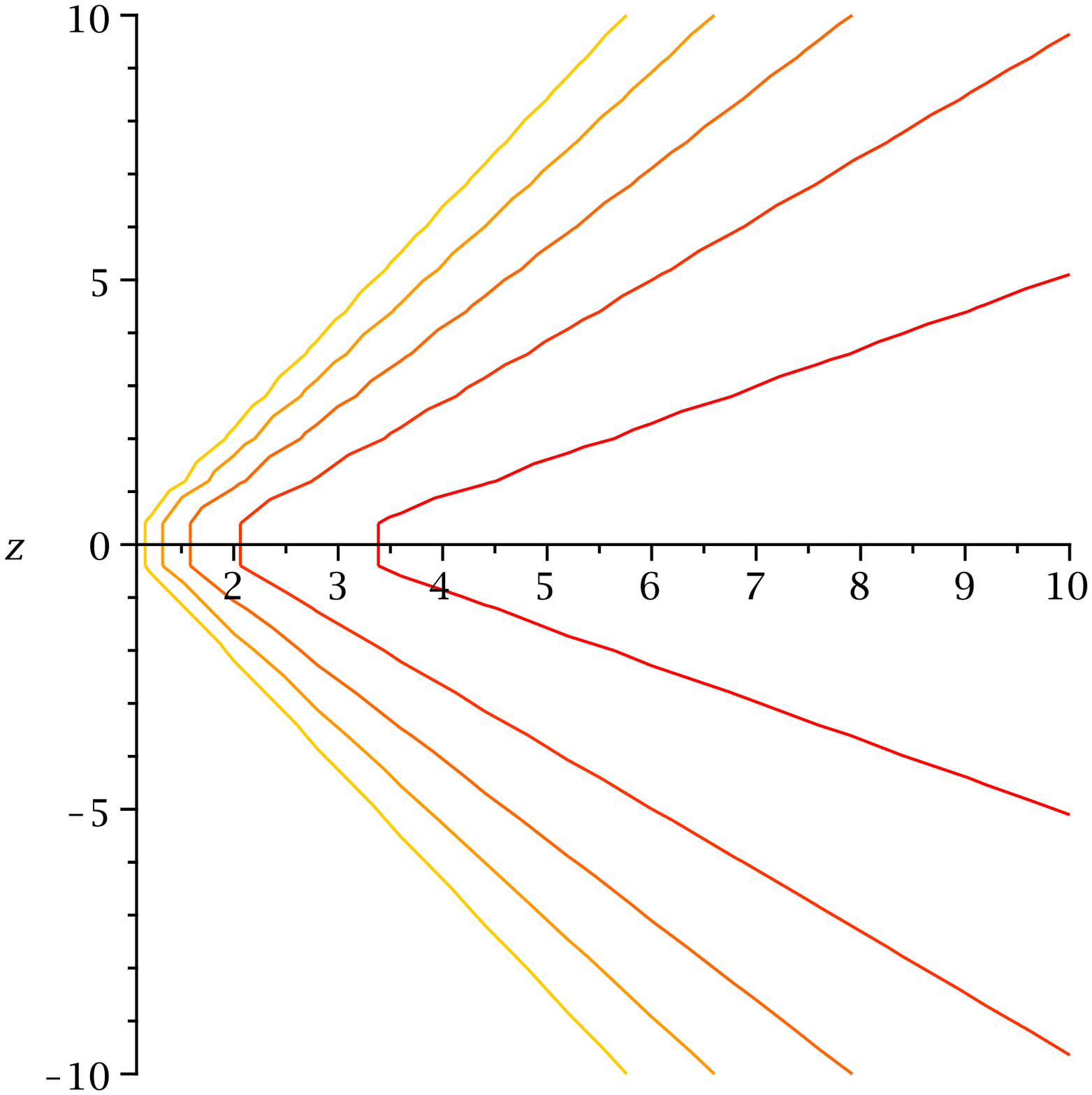} &
\includegraphics[width=0.3\textwidth]{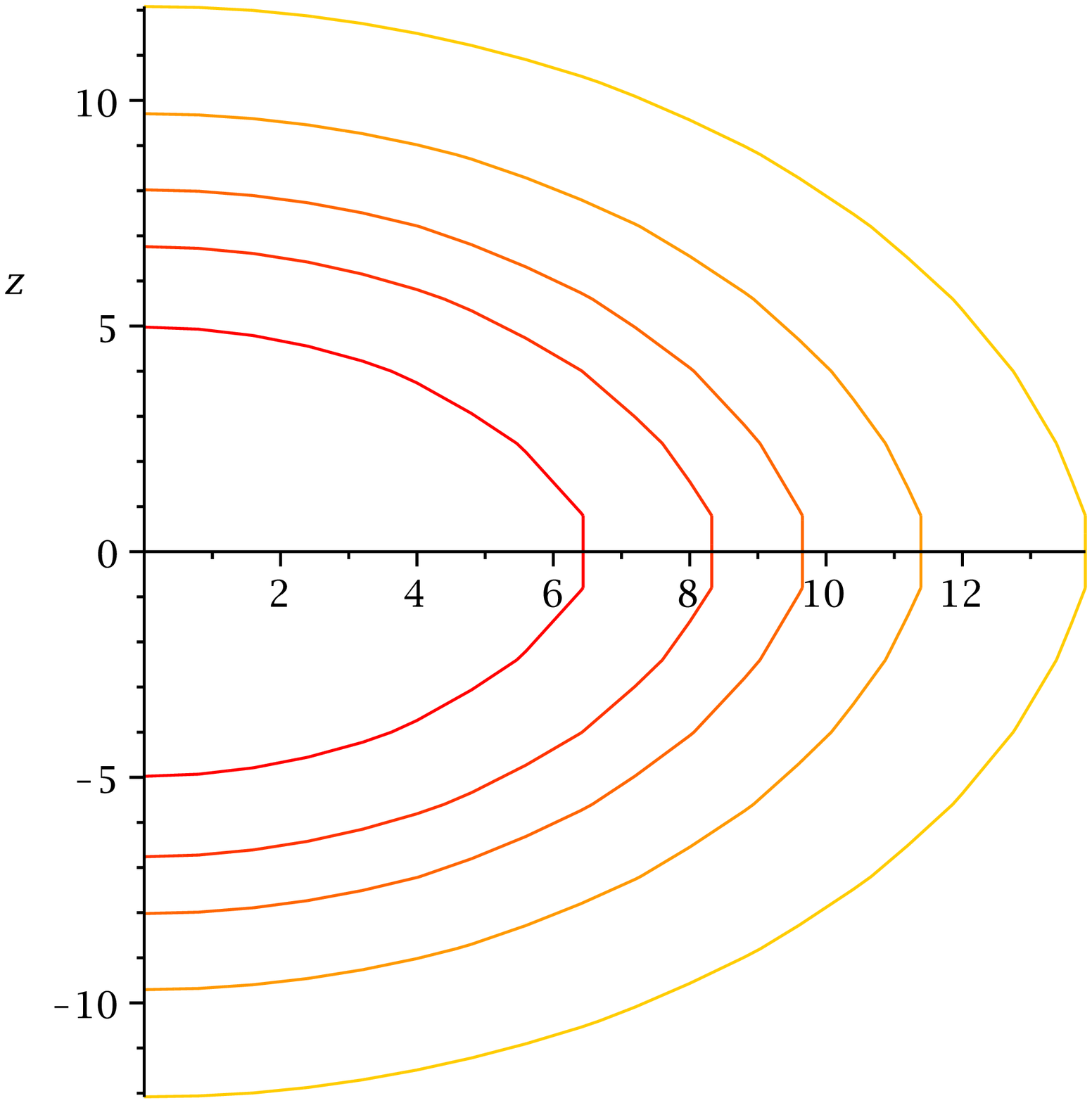}  \\
 \tilde R & \tilde R  \\
(a)    &  (b)
\end{array}
$$	
\caption{
Magnetic field lines   $(a)$  and  equipotential lines of the gravitational field 
$(b)$  for  $z_0=2$,  $b=1$,
$C_A = 2$, $3$, $4$, $5$, $6$,
$C_G= 0.9$, $0.92$, $0.93$, $0.94$, $0.95$.
Note that the gravitational field lines have the same behavior that  the magnetic one.  }
\label{fig:fig3}
\end{figure}

In Figs. \ref{fig:fig3}$(a)$ and \ref{fig:fig3}$(b)$,  we have drawn  the magnetic field lines  and 
equipotential lines of the gravitational field
for $z_0=2$, $b=1$, $C_A = 2$, $3$, $4$, $5$, $6$ ,
and $C_G= 0.9$, $0.92$, $0.93$, $0.94$, $0.95$.
Since gravitational field lines are perpendicular  to  equipotential  lines, we conclude that 
they have  a similar behavior   to the magnetic ones.

\section{Discussion}

A simple family  of  perfect fluid disks in presence of a magnetic field
was presented  based in a    magnetic Reissner-Nordstrom  solution.
The models were constructed using the well-known “displace, cut and reflect”
method.  The disks satisfy all the energy conditions.

We found  that the  the presence of  magnetic field
increases  the energy density  and the electric current density  while   the
pressure is lowered everywhere on the disks.
The same behavior was  observed for all the values of parameter.

We also analyzed the electrogeodesic equatorial circular motion of charged test particles around of the disks.
We found that  the presence of magnetic field provides the
possibility of to find relativist charged particles moving in both direct and retrograde
direction. We also observer that   the  circular speed  of particles is always a quantity
less than the speed  of light and that the magnetic field makes more relativistic these orbits.

Finally, other magnetostatic axially symmetric solutions that can be  written in conformastatic form
are  being investigated.

\begin{acknowledgments}
G. G. R.  wants to thank   the warm
hospitality of to Departamento de Gravitaci\'on
y Teor\'{\i}as ́de Campos (ICN-UNAM), where  this work was completed.
He also wants to thank
to  A. C. Guti\'errez-Pi\~neres for  useful comments and discussions.

\end{acknowledgments}

%\end{document}

\newpage

\end{document}